\pdfoutput=1
\documentclass{sf2a-conf2017}
\usepackage{amsmath}
\usepackage{graphicx}
\usepackage{hyperref}
\usepackage[]{natbib}  
\usepackage{epstopdf}

\usepackage{color}
\usepackage[normalem]{ ulem }
\usepackage{soul}

\def\BibTeX{{\rm B\kern-.05em{\sc i\kern-.025em b}\kern-.08em
    T\kern-.1667em\lower.7ex\hbox{E}\kern-.125emX}}
\bibpunct{(}{)}{;}{a}{}{,}  

\begin{document}

\TitreGlobal{SF2A 2017}


\title{Modeling Radial Velocities and Eclipse Photometry of the Kepler Target KIC\,4054905: an Oscillating Red Giant in an Eclipsing Binary}

\runningtitle{Red giants in eclipsing binaries}


\author{M. Benbakoura$^{1,}$}\address{IRFU, CEA, Universit\'e Paris-Saclay, F-91191 Gif-sur-Yvette, France}\address{Universit\'e Paris Diderot, AIM, Sorbonne Paris Cit\'e, CEA, CNRS, F-91191 Gif-sur-Yvette, France}

\author{P. Gaulme$^{3,4,}$}\address{Max-Planck-Institut f\"{u}r Sonnensystemforschung, Justus-von-Liebig-Weg 3, 37077, G\"{o}ttingen, Germany}\address{Department of Astronomy, New Mexico State University, P.O. Box 30001, MSC 4500, Las Cruces, NM 88003-8001, USA}\address{Physics Department, New Mexico Institute of Mining and Technology, 801 Leroy Place, Socorro, NM 87801, USA}

\author{J. McKeever$^{4,}$}\address{Department of Astronomy, Yale University, 52 Hillhouse Avenue, New Haven, CT 06511, USA }

\author{P. G. Beck$^{7,}$}\address{Instituto de Astrof\'{\i}sica de Canarias, E-38200, La Laguna, Tenerife, Spain}\address{Universidad de La Laguna, Dpto. de Astrof\'{\i}sica, E-38205, La Laguna, Tenerife, Spain}

\author{J. Jackiewicz$^{4}$}

\author{R. A. Garc\'{i}a$^{1,2}$}

\setcounter{page}{237}


\maketitle

\begin{abstract}
Asteroseismology is a powerful tool to measure the fundamental properties of stars and probe their interiors. This is particularly efficient for red giants because their modes are well detectable and give information on their deep layers. However, the seismic relations used to infer the mass and radius of a star have been calibrated on the Sun. Therefore, it is crucial to assess their accuracy for red giants which are not perfectly homologous to it. We study eclipsing binaries with a giant component to test their validity. We identified 16 systems for which we intend to compare the dynamical masses and radii obtained by combined photometry and spectroscopy to the values obtained from asteroseismology. In the present work, we illustrate our approach on a system from our sample.
\end{abstract}



\section{Introduction}
















Over the last 10 years, asteroseismology has been the most efficient tool to probe stellar interiors. In this field, red giants are of particular interest because, on the one hand, they are solar-like pulsators, i. e., their oscillations are stochastically excited by granulation, and on the other hand, the presence of mixed-modes in their power spectrum allows to probe their core \citep[e.g.][]{Becketal2011,Becketal2012,Mosseretal2011}. Moreover, the huge sample of observed red giants from \textit{Kepler} \citep{Boruckietal2010} and CoRoT \citep{Baglinetal2006}, including more than 30,000 stars, allows to make statistical inferences as a function of different parameters.

Obtaining masses and radii of red giants allows to know more about the fate of the Sun. These parameters can be directly obtained by measuring the effective temperature as ell as the global seismic parameters $\Delta \nu$ and $\nu_{\mathrm{max}}$, by applying the so-called global seismic scaling relations \citep{Brownetal1991,KjeldsenBedding1995}:
\begin{align}
\dfrac{R_*}{R_{\odot}} &= \left(\dfrac{\Delta \nu_{\odot}}{\Delta \nu}\right)^2 \left(\dfrac{\nu_{\mathrm{max}}}{\nu_{\mathrm{max},\odot}}\right) \left(\dfrac{T_{\mathrm{eff}}}{T_{\mathrm{eff},\odot}}\right)^{1/2} \\
\dfrac{M_*}{M_{\odot}} &= \left(\dfrac{\Delta \nu_{\odot}}{\Delta \nu}\right)^4 \left(\dfrac{\nu_{\mathrm{max}}}{\nu_{\mathrm{max},\odot}}\right)^4 \left(\dfrac{T_{\mathrm{eff}}}{T_{\mathrm{eff},\odot}}\right)^{3/2}
\end{align}

These relations have been calibrated on the Sun and are based on the assumption that solar-like pulsators are homologous to our star. However, in red giants the aspect ratio (radius of the radiative zone divided by the radius of the star) may be smaller than in main-sequence stars by more than one order of magnitude. Moreover, the envelope of the former is significantly less dense because of the expansion. The departure from homology and asymptotic regime might introduce a systematic error in the seismic relations \citep[e. g.][and references therein]{Mosseretal2013,Rodriguesetal2017} Therefore, the scaling relations should be studied carefully and it is crucial to test their accuracy.

One way of testing the validity of seismic scaling relations is to compare the values of stellar mass and radii from seismology with results from independent methods. \citet{Gaulmeetal2016} and \citet{Rawlsetal2016} considered eclipsing binary systems with at least one giant component. For such systems, the mass and radius of both components can be accurately measured by combining photometry and spectroscopy. Indeed, the simultaneous modeling of the eclipses and the radial velocities allows to retrieve, on the one hand, the orbital parameters of the systems such as eccentricity, inclination, longitude of periastron, and semi-major axis, and on the other hand, global properties of its components. On a sample of 10 systems, \citet{Gaulmeetal2016} compared the values of masses and radii obtained through this method (the dynamical values) with the seismic inferences. They found that the latter were systematically overestimated by $\sim 5 \%$ for the radius and $\sim 15\%$ for the mass.

To investigate further this result, we aim at extending this set of stars to a larger fraction of multiple systems containing at least one pulsating red giant. Hence, we have identified 16 new binary systems with a giant component. All of them have been observed by \textit{Kepler}. We combine the photometric light curves with spectroscopic follow-up observations from the ARCES spectrometer on the 3.5m telescope at Apache Point Observatory (APO), New Mexico, and the HERMES spectrometer \citep{Raskinetal2011} on the 1.2m Mercator Telescope in La Palma to measure the dynamical masses and radii and compare them to the seismic inferences. The latter were obtained from \textit{Kepler} data prepared with KADACS software \citep{Garciaetal2011}, which was analyzed by the asteroseimic pipeline A2Z \citep{Mathuretal2010}.

In this work, we show our preliminary analysis of KIC~4054905. Section~\ref{sec:dynamical_measurements} details how we measured the dynamical mass and radius of the giant. In section~\ref{sec:seismic_inferences}, we explain how we prepared the light curve from which the global seismic parameters were computed. Then, we show the result of the comparison between seismic estimates and dynamical measurements. Finally, in section~\ref{sec:conclusion}, we present the perspectives of our work and the suggestions to improve the scaling relations.

\section{Measuring the dynamical masses and radii}
\label{sec:dynamical_measurements}











The photometric light curves were obtained from the Target Pixel Files, which we integrated to produce time series. They were then flattened so that only the eclipses remain. We modeled the orbit with the JKTEBOP code \citep{Southworth2013}, which uses a Markov chain Monte Carlo (MCMC) method to fit the sum of the fractional radii $\frac{R_1 + R_2}{a}$, where $R_1$, $R_2$ and $a$ are the radius of the giant, that of the companion and the semi-major axis, respectively, the ratio of the radii $\frac{R_2}{R_1}$, the orbital inclination and eccentricity, the longitude of periastron, the brightness ratio $\left(\frac{T_{\mathrm{eff},2}}{T_{\mathrm{eff},1}}\right)^4$, orbital period and reference time for the primary eclipse. Concerning the eclipses, we used the same convention as \citet{Gaulmeetal2016}, i.e., the primary eclipse refers to the companion eclipsing the giant. We set its orbital phase to 0. For the limb darkening of the companion, we assumed a linear law. We did not fit it because of the small influence it has compared to that of the giant and the resulting difficulty to converge on this parameter. Concerning the giant, we assumed a quadratic law and computed guesses with the JKTLD routine, which uses limb-darkening tables to interpolate them in the \textit{Kepler} bandpass. We fixed the order-2 coefficient and fitted the linear one.

\begin{figure}
\centering
\includegraphics[width=.8\linewidth]{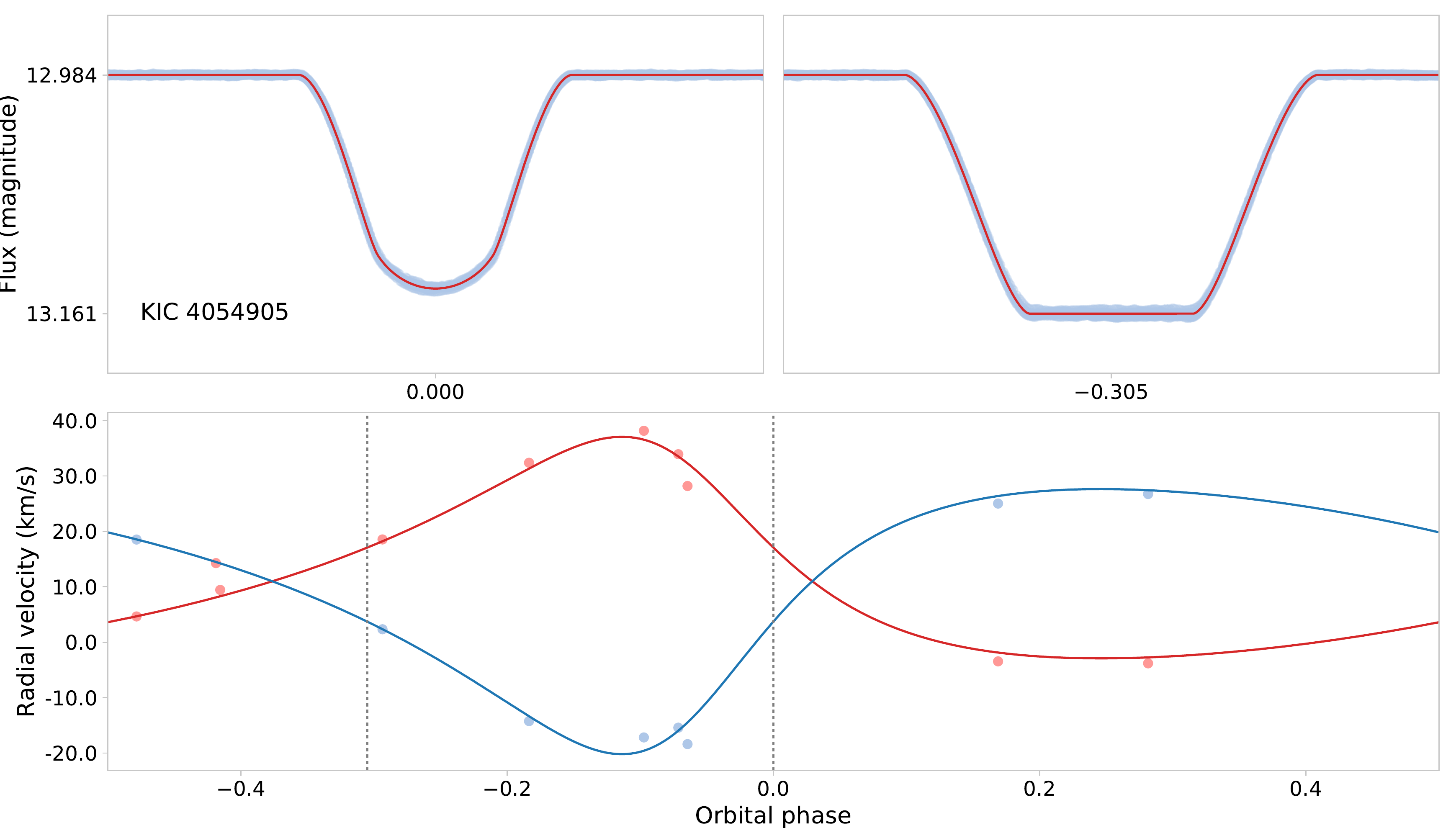}
\caption{Eclipses and radial velocities of KIC~4054905. \textit{Top left panel:} Zoom on the primary eclipse (companion eclipsing the giant. The blue dots represent the processed phase-folded \textit{Kepler} light curve and the red line, the fit of JKTEBOP. \textit{Top right panel:} zoom on the secondary eclipse (giant eclipsing the companion). \textit{Bottom panel:} Radial velocities of the components of the systems. The red and blue dots represent the measurements done for the giant and the companion at APO, respectively. The red and blue solid lines represent the fit of JKTEBOP for each component, respectively. The vertical dotted lines the orbital phases corresponding to the primary and second eclipse.}
\label{fig:4054905_eclipses_RVs}
\end{figure}

We combined the modeling of the eclipses with a two-year spectroscopic follow-up done mostly at APO. We used the Echelle spectrometer ARCES located at the 3.5m telescope. The data was reduced as described by \citet{Rawlsetal2016}. We used the broadening function (BF) technique as outlined by \citet{Rucinski2002} to extract the radial velocities. This method relies on solving a convolution equation and requires a template spectrum to compare the positions of the spectral lines. In this work, we used templates from the PHOENIX stellar atmosphere model \citet{Husseretal2013}. Since the giant is in general three to ten times more luminous than its companion, its lines are significantly easier to detect. This is why we used a template of a main-sequence star to compute the BF. Its temperature was chosen the closest to that of the companion. Combining radial velocities to the eclipses allows to break the degeneracy in the model. Thus, it is possible to obtain the semi-major axis of the system and the separate masses and radii of each component. Figure~\ref{fig:4054905_eclipses_RVs} illustrates the modeling of combined eclipses and radial velocities of KIC~4054905. For this star, the dynamical analysis gives $M_{\mathrm{giant}} = 1.05\ M_{\odot}$ and $R_{\mathrm{giant}} = 8.24\ R_{\odot}$.

\section{Seismic inferences}
\label{sec:seismic_inferences}







We prepared the light curves from the \textit{Kepler} Target Pixel Files as described by \citet{Garciaetal2011}. In this procedure, we inserted an intermediate step to remove the eclipses from the signal. To that end, we used the Eclipsing Binary Catalog, which provides information on eclipses timing and durations \citep{Abdul-Masihetal2016}. Removing the eclipses in the time series generates gaps in the data. To eliminate their signature, we inpainted the light curves \citep{Garciaetal2014inpaint,Pires2015}. From the obtained time series, we computed the power spectra and used the pipeline A2Z to compute the global seismic parameters $\Delta \nu$ and $\nu_{\mathrm{max}}$. For KIC~4054905, we obtained $M_{\mathrm{seismo}} = 1.3 \pm 0.1 \ M_{\odot}$ and $R_{\mathrm{seismo}} = 9.3 \pm 0.7 \ R_{\odot}$.


\begin{figure}
\centering
\includegraphics[width=.8\linewidth]{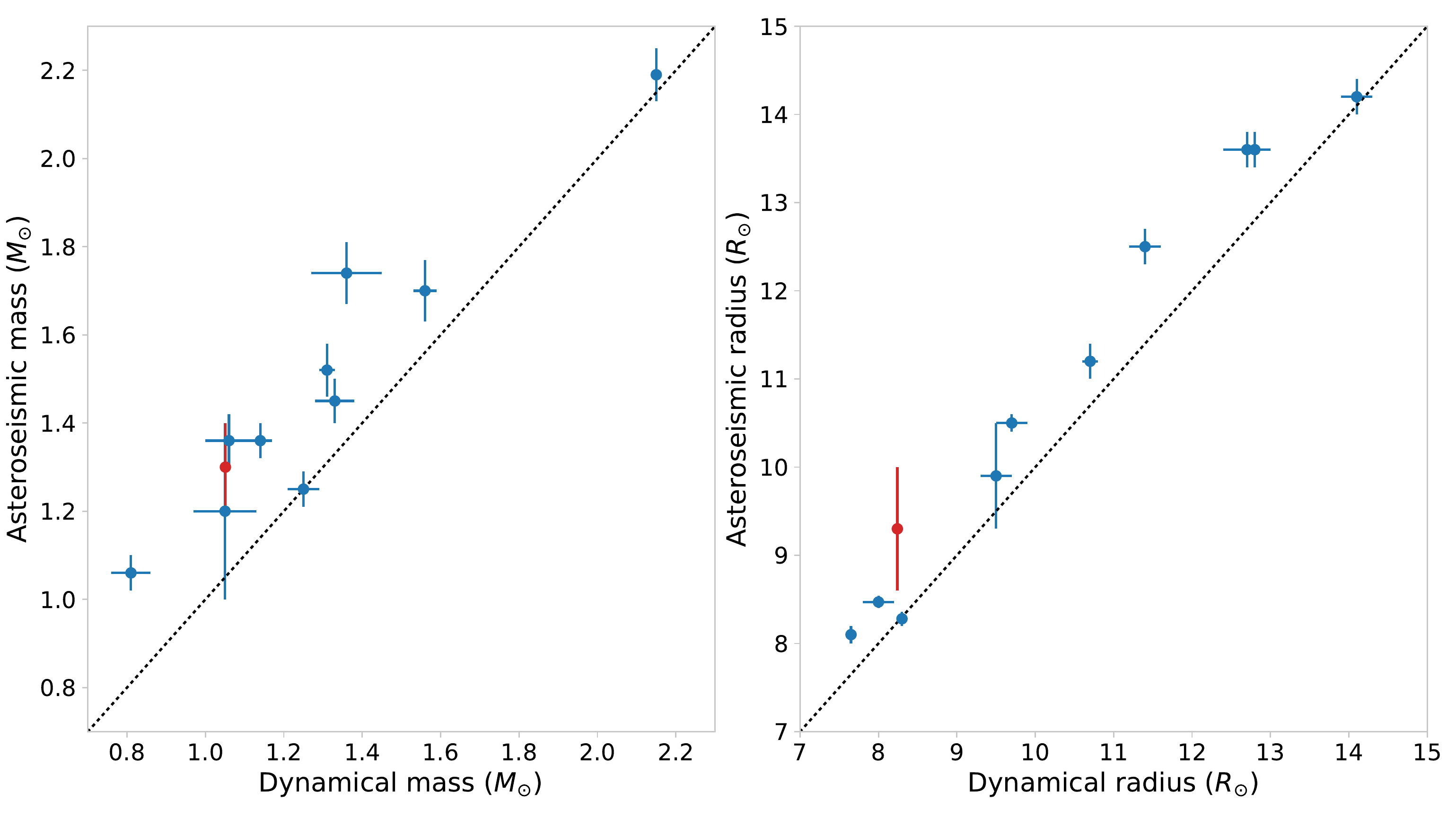}
\caption{Comparison between dynamical measurements and seismic inferences of masses and radii. \textit{Left panel:} Seismic masses as a function of dynamical masses. The blue dots correspond to the sample of \citet{Gaulmeetal2016} and the red dot, to our measurements on KIC\,4054905. The grey dotted line represents the line on which $M_{\mathrm{seismo}} = M_{\mathrm{dyn}}$. \textit{Right panel:} Seismic radii as a function of dynamical radii.}
\label{fig:compare_dyn_ast}
\end{figure}

Figure~\ref{fig:compare_dyn_ast} presents the comparison between dynamical measurements and seismic estimates of masses and radii for the sample of \citet{Gaulmeetal2016} and KIC\,4054905. Our preliminary result is in good agreement with their study since both mass and radius of our star are over estimated by the scaling relations.

\section{Conclusion}
\label{sec:conclusion}




Our preliminary result is in agreement with those of \citet{Gaulmeetal2016}, suggesting that the scaling relations systematically overestimate mass and radius. However, more spectroscopic observations of the other targets of our sample are needed to draw more robust conclusions. Extending the sample is crucial to test their accuracy. The departure of red giants from the asymptotic regime was already discussed by \citet{Mosseretal2013}. Recently, \citet{Rodriguesetal2017} proposed to improve the scaling relations by taking into account that red giants depart from homology to the Sun. Other reasons for discrepancy between dynamical measurements and seismic estimates should also be investigated, such as the effect of other observables. For instance, \citet{Corsaroetal2017} found  of metallicity influencing granulation in stellar clusters. Thus, considering new observables is important to better understand stellar dynamics and asteroseismology.

%
%
%
%
%
%
%
%

%

\newpage

\begin{acknowledgements}
We acknowledge the work of the team behind \textit{Kepler}. Funding for the \textit{Kepler} Mission is provided by NASA's Science Mission Directorate. 
We thank the technical team as well as the observers of the \textsc{Hermes} spectrograph and Mercator Telescope, operated on the island of La Palma by the Flemish Community, at the Spanish Observatorio del Roque de los Muchachos of the Instituto de Astrof{\'i}sica de Canarias. 
The authors acknowledge S. Mathur for useful comments and discussion during the project. M.B. and P.G. acknowledge support from NASA ADAP grant
NNX14AR85G. M.B. and R.A.G. received funding from the CNES. PGB acknowledges the support of the Spanish Ministry of Economy and Competitiveness (MINECO) under the programme 'Juan de la Cierva' (IJCI-2015-26034). 
\end{acknowledgements}

\bibliographystyle{aa}  
\bibliography{sf2a-template} 

\end{document}